# Engineering Ferroelectric $Hf_{0.5}Zr_{0.5}O_2$ Thin Films by Epitaxial Stress

Saúl Estandía,[1] Nico Dix,[1] Jaume Gazquez,[1] Ignasi Fina,[1] Jike Lyu,[1] Matthew F. Chisholm,[2] Josep Fontcuberta,[1] Florencio Sánchez[1,*]

[1] Institut de Ciència de Materials de Barcelona (ICMAB-CSIC), Campus UAB, Bellaterra 08193, Barcelona, Spain

[2] Center for Nanophase Materials Sciences, Oak Ridge National Laboratory, Tennessee 37831-6064, USA

ABSTRACT: The critical impact of epitaxial stress on the stabilization of the ferroelectric orthorhombic phase of hafnia is proved. Epitaxial bilayers of $Hf_{0.5}Zr_{0.5}O_2$ (HZO) and $La_{0.67}Sr_{0.33}MnO_3$ (LSMO) electrodes were grown on a set of single crystalline oxide (001)-oriented (cubic or pseudocubic setting) substrates with lattice parameter in the 3.71 – 4.21 Å range. The lattice strain of the LSMO electrode, determined by the lattice mismatch with the substrate, is critical in the stabilization of the orthorhombic phase of HZO. On LSMO electrodes tensile strained most of the HZO film is orthorhombic, whereas the monoclinic phase is favored when LSMO is relaxed or compressively strained. Therefore, the HZO films on $TbScO_3$ and $GdScO_3$ substrates present substantially enhanced ferroelectric polarization in comparison to films on other substrates, including the commonly used $SrTiO_3$. The capability of having epitaxial doped $HfO_2$ films with controlled phase and polarization is of major interest for a better understanding of the ferroelectric properties and paves the way for fabrication of ferroelectric devices based on nanometric $HfO_2$ films.

1. INTRODUCTION

The recent demonstration of ferroelectricity in nanometric thin films of a metastable orthorhombic phase of doped $HfO_2$ [1] opens promising opportunities for memory devices[2-4] and energy applications.[2,5-7] The metastable phase of $HfO_2$ is usually crystallized by annealing thin film heterostructures of amorphous hafnia sandwiched between top and bottom electrodes, typically TiN [2,8-9] or TaN.[10] The resulting films are polycrystalline and contain paraelectric tetragonal and monoclinic phases besides the ferroelectric orthorhombic phase.[1,2,11-12] The ferroelectric phase has been also grown epitaxially on a few substrates, including yttria-stabilized zirconia,[13-16] $LaAlO_3$,[17] $SrTiO_3$,[18-21] and buffered Si.[22] The research on epitaxial stabilization is just emerging in comparison with that on polycrystalline doped $HfO_2$ films.[2,5,8,10-12,23] However, epitaxial $HfO_2$ films are of huge interest as their properties can be better controlled than those of polycrystalline samples. Besides the single crystal orientation in epitaxial films, the control of the epitaxial stress can permit engineering of the microstructure and the resulting ferroelectric properties of the films. The relevance of epitaxial stress on the growth of ferroelectric $HfO_2$ is two-fold. On one hand, epitaxial stress affects greatly the energy of a (semi)coherent interface between a substrate and a heteroepitaxial film, and it can favor the stabilization of a metastable phase that is in competition with other polymorphs. This epitaxial stabilization has been used to obtain unstable phases of a variety of complex oxides.[24-27] On the other hand, epitaxial stress can cause elastic lattice strain, which can modify the energy of the polymorphs[23,28,29] and can also produce important effects on the polarization of ferroelectric oxides.[30-31] The most common method to control stress in heteroepitaxial films is based on the selection of a substrate with particular lattice mismatch. However, this substrate engineering remains unexplored for ferroelectric $HfO_2$.

Aiming to investigate the effects of epitaxial stress, epitaxial $Hf_{0.5}Zr_{0.5}O_2$ (HZO) films were grown on a set of single crystalline oxide substrates presenting a wide range of lattice parameters (**Figure 1**a). $La_{0.67}Sr_{0.33}MnO_3$ (LSMO) epitaxial electrodes and HZO films were sequentially deposited in a single process. The lattice parameter of LSMO is expected to be critical on the epitaxial stabilization of HZO since the electrode is the epitaxial template on which HZO grows. It is found that the substrate determines the lattice strain of LSMO, and that the LSMO strain state strongly influences the formation of orthorhombic and monoclinic phases in the HZO film.

Therefore, the substrate determines the amount of orthorhombic phase, and substrate selection permits tuning of the ferroelectric polarization of the film. The remnant polarization $P_r$ ranges from less than 5 μC/cm$^2$ for films on LSAT ($a_s$ = 3.868 Å) and substrates with smaller lattice parameter to around 25 μC/cm$^2$ for films on $TbScO_3$ ($a_s$ = 3.96 Å) and $GdScO_3$ ($a_s$ = 3.97 Å). The results demonstrate that tensile strained LSMO electrodes favor the epitaxial stabilization of the metastable orthorhombic phase and enhancement of the ferroelectric polarization. Therefore, epitaxial stress engineering can be successfully applied to $HfO_2$, allowing control of the ferroelectric properties and making possible increased polarization. This control, which does not require varying thickness or deposition parameters, can pave the way to understand the correlations between structural and ferroelectric properties of $HfO_2$, and it is relevant for prototyping devices based on nanometric $HfO_2$ films.

## 2. RESULTS AND DISCUSSION

**Figure 1**b shows X-ray diffraction (XRD) θ-2θ symmetric scans of the HZO/LSMO/substrate samples. The position of the LSMO 002 reflection, in the 2θ = 44 - 49° range, depends strongly on the substrate. The corresponding LSMO out-of-plane (oop) lattice parameters are presented in **Figure 1**c as a function of the lattice mismatch with the substrate. The in-plane (ip) lattice parameters, determined from reciprocal space maps (RSM) around asymmetrical reflections (**Figure S1**), are also shown. Increasing lattice mismatch from around -2% to 2% (see the zoom in **Figure 1**d), reduces the LSMO oop parameter monotonously from ~4.00 Å (on $LaAlO_3$) to ~3.80 Å (on $TbScO_3$), whereas the corresponding ip parameter increases to ~3.79 Å to ~3.95 Å. The RSMs in **Figure S1** confirm that LSMO films, 25 nm thick, are elastically strained in the -2 to +2% lattice mismatch range. In the films on substrates having larger negative or positive lattice mismatch, the LSMO relaxes plastically and the oop and ip parameters approach the bulk value, almost matching it on the greatly mismatched (<-4.1%) $YAlO_3$ and (>8%) MgO substrates. Therefore, it is proven that the strain state of LSMO is determined by the substrate.

The LSMO electrodes act as a template for the subsequent growth of HZO, and thus their lattice strain, that depends on the substrate, can be relevant to the epitaxy of HZO. The XRD θ-2θ scans in **Figure 1**b show orthorhombic (o) HZO(111) at 2θ ~30° and/or monoclinic (m) HZO(002) at 2θ ~34° diffraction peaks. The m-HZO peak can include contribution of 200, 020 and 002 reflections. The peaks are broad due to the nanometric thickness of the layers (~9.5 nm).

The amount of each phase depends on the substrate. In the case of the substrates with lattice parameter from 3.905 Å ($SrTiO_3$) to 4.01 Å ($NdScO_3$), the o-HZO(111) peaks have high intensity and Laue fringes can be seen (**Figure S2**). The XRD 2θ-χ frames of all samples, presented in **Figure 2**a, reflect the impact of the substrate on the HZO phases. It is noticed that the o-HZO 111 is a bright circular spot, whereas the m-HZO 002 reflection is generally elongated along χ, signaling higher mosaicity (excluding the film on $LaAlO_3$, which m-HZO 002 reflection is a bright spot). To map the formation of orthorhombic and monoclinic phases as a function of the substrate, the intensity at each 2θ was integrated from χ = -10° to χ = +10° for each frame (**Figure 2**b). The intensity is plotted in a logarithmic color scale and the 2θ scans are shifted vertically, ordered as the lattice parameter of the substrate increases (see labels at the right). The 001 reflections of substrate (marked with red dashed line) and LSMO electrode (marked with black dotted line) are at 2θ angles from around 20° to 26°, and the corresponding 002 reflections are from around 40° to 50°. The o-HZO 111 and the m-HZO 002 reflections are at around 30° and 34°, respectively. The map shows that the orthorhombic phase is mainly present on substrates with lattice parameter from 3.905 Å ($SrTiO_3$) to 4.01 Å ($NdScO_3$), and that basically pure orthorhombic phase films only are obtained on $DyScO_3$, $TbScO_3$, and $GdScO_3$. **Figure 2**b indicates that the orthorhombic phase is favored on substrates with large lattice parameter, whereas the amount of monoclinic phase is greater when the lattice parameter of the substrate is smaller. The films on substrates with very large or very small lattice parameter do not follow this tendency, which is likely due to the plastic relaxation of the LSMO electrode. Indeed, the plot of the intensity of the reflections of both phases shows monotonic dependences on the ip lattice parameter of LSMO (**Figure 2**c). The orthorhombic phase forms when the ip parameter of the LSMO template is larger than around 3.87 Å, and the XRD o-HZO 111 spot intensity increases with the LSMO ip parameter. The monoclinic phase shows an opposite tendency, and it is only absent when the LSMO template has an elongated ip parameter around 3.95 Å. The intensity of orthorhombic and monoclinic XRD reflections depends strongly on the ip parameter of the LSMO electrode (**Figure 2**c). It shows that depending on the strain of the LSMO electrode, either pure monoclinic phase, mixture of both phases, or pure orthorhombic films are obtained. In contrast, the lattice parameter of the LSMO template has little influence on the interplanar $d_{o\text{-}HZO(111)}$ spacing (**Figure 2**d), and only a slight $d_{o\text{-}HZO(111)}$ contraction, close to the detection limit, can be appreciated in the films on electrodes with largest $a_{LSMO}$. This suggests plastic relaxation, which is confirmed by XRD reciprocal space maps (**Figure S3**).

Polymorphs that are unstable in bulk materials can form in thin films due to the change in energy in case of elastic strain and the contribution of surface and interface energies. Density functional (DFT) calculations[23,28,29] predict for $HfO_2$ that compressive strain and surface energy contribution reduce the energy of the polar orthorhombic phase with respect to the monoclinic one. Thus, the formation of the polar phase is more favored in ultrathin films where the surface energy contribution is more relevant. These DFT calculations considered films having {100} [23,28] and {110} [32] orientations. Very recently, DFT calculations were extended to (111) orientation, and remarkably it was found that the orthorhombic phase in (111)-oriented films has minimum energy for positive strain around 1.5%, and its energy was smaller than that of (111)-oriented monoclinic phase for a very broad range of strain extending from negative values to positive values well above 2%.[29] Therefore, the o-HZO(111) orientation in our epitaxial films can be a relevant factor on the stabilization of the ferroelectric phase, although its formation in our films is competing with the {100} orientation of the monoclinic phase. On the other hand, strain is likely less relevant considering the low elastic strain (Figure S3) of the films. In addition, the interface between HZO and the bottom surface (the LSMO electrode in this case), for which energy calculations are not reported, can be determinant on the total energy of HZO polymorphs. The epitaxial stabilization of o-HZO with (111)-orientation implies a change in crystal symmetry, being the HZO film (111) oriented on the 4-fold symmetry LSMO(001) surface. Heteroepitaxy with different symmetry between a top layer and a bottom layer (or the substrate) is relatively frequent.[33] Films can present either higher[34] or lower[35] symmetry than the substrate. Epitaxy requires matching between layer and substrate crystal lattices, which is intriguing when the surface symmetry of layer and substrate is different. However, heteroepitaxy can happen in largely mismatched film-substrate systems by coincidence of m lattice planes of the film on n planes of the substrate.[36] This mechanism is often observed in heteroepitaxy of semiconductors[36] and oxides[37]. The change in symmetry usually causes formation of crystal variants, like in the case of o-HZO(111) films on LSMO(001) surfaces. Related examples are epitaxial growth of spinel $NiFe_2O_4$(111) films on yttria-stabilized zirconia-YSZ(001)[38] or wurtzite ZnO(0001) on MgO(001)[35].

XRD pole figures around asymmetrical o-HZO -111 and m-HZO -111 reflections in **Figure 3**a, confirm that both orthorhombic and monoclinic phases, when present in the films, are epitaxial (see ϕ-scans in **Figure S4**). The sample on $GdScO_3$ shows 4 sets of three high intensity o-HZO -111 spots, indicating the existence of four crystal variants with 90° rotation in the plane. In **Figure 3**c the epitaxial relationship is sketched. The rhombohedral distortion reported[20] in similar HZO films on LSMO(001) electrodes is not observed here within the sensitivity of the XRD measurements. In the films on substrates with smaller lattice parameter, $SrTiO_3$ and LSAT, the intensity of the o-HZO -111 reflections decreases, and they are not observed in the film on $LaAlO_3$. In contrast, the poles around m-HZO -111, show four intense spots in the film on $LaAlO_3$, lower intensity spots on LSAT and $SrTiO_3$, and barely detectable on $GdScO_3$. The epitaxial relationship of this phase is sketched in **Figure 3**d.

Topographic atomic force microscopy (AFM) images of HZO/LSMO bilayers on $NdScO_3$, $DyScO_3$, LSAT, and $LaAlO_3$ are shown in **Figures 4** (a-d), respectively. The films are very flat, with root means square (rms) roughness less than 5 Å, and a morphology of terraces and steps can be appreciated in some of the images. The surface flatness of all the films is remarkable considering the broad range of lattice parameter of the substrates. In **Figure S5** we show AFM images of all the samples and the rms roughness is plotted as a function of the lattice parameter of the substrate. It is seen that roughness increases from around 2 Å to 4 - 5 Å with the substrate lattice parameter.

Scanning transmission electron microscopy (STEM) has been used for structural characterization at the nanoscale and to identify the orthorhombic and monoclinic phases in HZO films on three substrates with fully strained LSMO electrodes: i) LSAT, with lattice parameter ($a_s$ = 3.868 Å) and the best lattice matching with bottom LSMO electrode, ii) $SrTiO_3$, with larger lattice parameter $a_s$ = 3.905 Å, which has been already used[18-20] for epitaxial growth of o-HZO(111), and iii) $GdScO_3$, with much larger lattice parameter ($a_s$ = 3.97 Å). The corresponding cross-sectional high-angle annular dark field (HAADF) images are presented in **Figures 5**a-c, respectively. The low magnification (top panels) and high magnification (bottom panels) images were obtained along the [110] zone axes of the substrates. The low magnification images show a clear contrast between the HZO film, LSMO electrode, and substrate. In order to properly identify the phases, their orientation and epitaxy, HAADF images are compared with the orthorhombic and monoclinic projected structures. High magnification images of the HZO film on LSAT (**Figure 5**a, bottom panel) confirm the presence of the monoclinic phase (see the inset), and the absence of the orthorhombic phase in the imaged section. In contrast, orthorhombic and monoclinic HZO crystallites coexist in the film on $SrTiO_3$. Insets in **Figure 5**b, bottom panel, show enlarged views of the monoclinic and the orthorhombic grains, with the projected structures superimposed. The lateral size of orthorhombic grains is around 10 ± 4 nm, while a slightly larger lateral size around 15 ± 5 nm is found for the monoclinic phase. Finally, the HZO film on $GdScO_3$ only presents orthorhombic grains (see the inset in bottom panel of **Figure 5**c), with absence of monoclinic phase in the imaged section. The epitaxial relationship for the orthorhombic phase is [1-10]o-HZO(111)//[1-10]LSMO(001)//[1-10]Substrate(001), where all the indices refer to the cubic or pseudocubic unit cells. For the m-phase, the epitaxial relationship is [010]m-HZO(001)//[1-10]LSMO(001)//[1-10]Substrate(001). These results demonstrate the huge impact of the substrate lattice parameter in the formation of monoclinic or orthorhombic HZO phase.

Ferroelectric polarization loops of the HZO films deposited on the different substrates are shown in **Figure 6**a. Detailed information about the ferroelectric measurement is presented in **Figure S6**. In agreement with the critical role of the substrate on the stabilization of the orthorhombic phase, the ferroelectricity is found to depend strongly on the substrate. The HZO films on the substrates with smaller lattice parameter, $YAlO_3$, $LaAlO_3$ and $NdGaO_3$, have low ferroelectric polarization of about 4 $\mu C/cm^2$. HZO films on substrates having larger lattice parameter show an increasing remnant polarization ($P_r$) from around 5 $\mu C/cm^2$ on LSAT to around 24 $\mu C/cm^2$ on $TbScO_3$. With further increase of the lattice parameter of the substrate, the polarization of HZO becomes progressively smaller, getting reduced to 9 $\mu C/cm^2$ in the film on MgO. The remnant polarization is plotted against the substrate lattice parameter in **Figure 6**b, showing a peaked dependence with largest polarization for HZO films on scandates with lattice parameter around 3.96 Å. It should be noted that the HZO films do not grow directly on the substrate but on the LSMO bottom electrode that is fully strained only on substrates with lattice parameter in the 3.79 – 3.97 Å range. Indeed, the plot of the remnant polarization against the ip parameter of the LSMO electrode shows very low polarization when $a_{LSMO}$ is smaller than around 3.87 Å, and continuous linear increase for larger $a_{LSMO}$ parameter (**Figure 6**c).

Two potential contributions to the ferroelectric polarization can be considered. First, the amount of orthorhombic phase formed, and second the strain state of the resulting o-HZO(111) phase (**Figures 1-3**). These two contributions can ultimately determine the ferroelectric polarization. Therefore, the remnant polarization is plotted as a function of the interplanar $d_{o\text{-}HZO(111)}$ spacing (**Figure** 7a) and the normalized intensity of the XRD o-HZO 111 reflection (**Figure** 7b). The films with shorter interplanar $d_{o\text{-}HZO(111)}$ spacing appear to have larger polarization, but the graph does not show a clear dependence as error bars in lattice parameter are comparable to its variation. In contrast, **Figure** 7b clearly confirms that samples with the largest amount of orthorhombic phase (mainly on scandate substrates) also have the largest polarization. Thus, the role of epitaxial stress is unraveled: 1) it conditions the epitaxial stabilization of the orthorhombic phase, and 2) the amount of this phase determines the ferroelectric polarization. The impact is critical and films on scandate substrates present greatly enhanced ferroelectric properties.

## 3. CONCLUSIONS

In conclusion, the role of epitaxial stress on the stabilization of the metastable orthorhombic phase of HZO has been unraveled. LSMO bottom electrodes are elastically strained in a range of lattice mismatch from around -2 to 2 %, and thus HZO films can be integrated in capacitor heterostructures with a broad range of epitaxial stress by selection of the substrate. The amount of stabilized orthorhombic phase is enhanced on substrates with pseudocubic lattice parameter larger than around 3.87 Å. The orthorhombic HZO phase becomes strongly favored with respect to the monoclinic HZO phase for increasing substrate lattice parameters, as long as the fully strained state of the LSMO is maintained. $TbScO_3$ and $GdScO_3$ are the optimal substrates to stabilize the orthorhombic HZO phase, with negligible amount of paraelectric phase and much higher polarization than that of films on $SrTiO_3$ or LSAT substrates. Epitaxial ultrathin HZO films with enhanced properties on $TbScO_3$ and $GdScO_3$ substrates could be used in emerging devices such as ferroelectric tunnel junctions, with superior performance than epitaxial films on $SrTiO_3$(001).

## EXPERIMENTAL SECTION

**Thin films deposition**: Epitaxial bilayers formed by top HZO films and bottom LSMO bottom electrodes (t = 25 nm) were grown in a single process by pulsed laser deposition (KrF excimer laser). A set of ten (001)-oriented cubic and (110)-oriented rhombohedral and orthorhombic substrates were used. For the sake of simplicity, pseudocubic cell is used here for the rhombohedral and orthorhombic substrates, being their orientation (001) in this setting. The lattice (cubic or pseudocubic) parameter of the used substrates is in the $a_s$ = 3.71 –4.21 Å range (**Figure 1**a). The HZO films, 9.5 nm thick, were deposited at substrate temperature of 800 °C under dynamical oxygen pressure of 0.1 mbar. Additional information on growth conditions of HZO and LSMO is reported elsewhere.[18-19]

**Structural characterization**: The crystal structure (crystal phases of HZO and lattice parameters of LSMO and HZO) was characterized by X-ray diffraction using Cu Kα radiation. A Siemens D5000 diffractometer with point detector was used to measure symmetric 2θ scans. A Bruker D8, equipped with 2d detector Vantec 500, was used to acquire 2θ-χ frames and pole figures around o-HZO -111 and m-HZO -111 asymmetric reflections. Atomic force microscopy in dynamic mode was used to characterize surface topography. Microstructural characterization of selected samples was done by scanning transmission electron microscopy using a Nion UltraSTEM 200, operated at 200 kV and equipped with a 5th order Nion aberration corrector, and a JEOL ARM 200CF STEM with a cold field emission source (equipped with a CEOS aberration corrector). High-angle annular dark field images of cross-sectional specimens were recorded along the [110] zone axes of the substrates.

**Ferroelectric characterization**: Capacitor structures were obtained by *ex-situ* deposition through stencil masks of top platinum electrodes, 20 nm in thickness and 19 μm in diameter, by dc magnetron sputtering. Ferroelectric polarization loops were obtained at room temperature in top-bottom configuration by means of an AixACCT TFAnalyser2000 platform. Leakage contribution was compensated using dynamic leakage current compensation (DLCC) standard procedure.[39-40]

ASSOCIATED CONTENT

**Supporting Information**. XRD reciprocal space maps around asymmetric LSMO reflections. Simulation of Laue fringes around o-HZO 111 reflection. XRD reciprocal space maps around asymmetric HZO reflections. XRD ϕ-scans around asymmetrical reflections of o- and m-HZO phases. Topographic atomic force microscopy images of all films. Information about measurement of ferroelectric polarization loops, including current-voltage curves

AUTHOR INFORMATION

**Corresponding Author**

* E-mail: fsanchez@icmab.es

ACKNOWLEDGMENTS

Financial support from the Spanish Ministry of Economy, Competitiveness and Universities, through the “Severo Ochoa” Programme for Centres of Excellence in R&D (SEV-2015-0496) and the MAT2017-85232-R (AEI/FEDER, EU), and MAT2015-73839-JIN projects, and from Generalitat de Catalunya (2017 SGR 1377) is acknowledged. IF and JG acknowledge Ramón y Cajal contracts RYC-2017-22531 and RYC-2012-11709, respectively. SE acknowledges the Spanish Ministry of Economy, Competitiveness and Universities for his PhD contract (SEV-2015-0496-16-3) and its cofunding by the ESF. JL is financially supported by China Scholarship Council (CSC) with No. 201506080019. SE and JL work has been done as a part of their Ph.D. program in Materials Science at Universitat Autònoma de Barcelona. The electron microscopy performed at ORNL was supported by the Materials Sciences and Engineering Division of Basic Energy Sciences of the Office of Science of the U.S. Department of Energy.

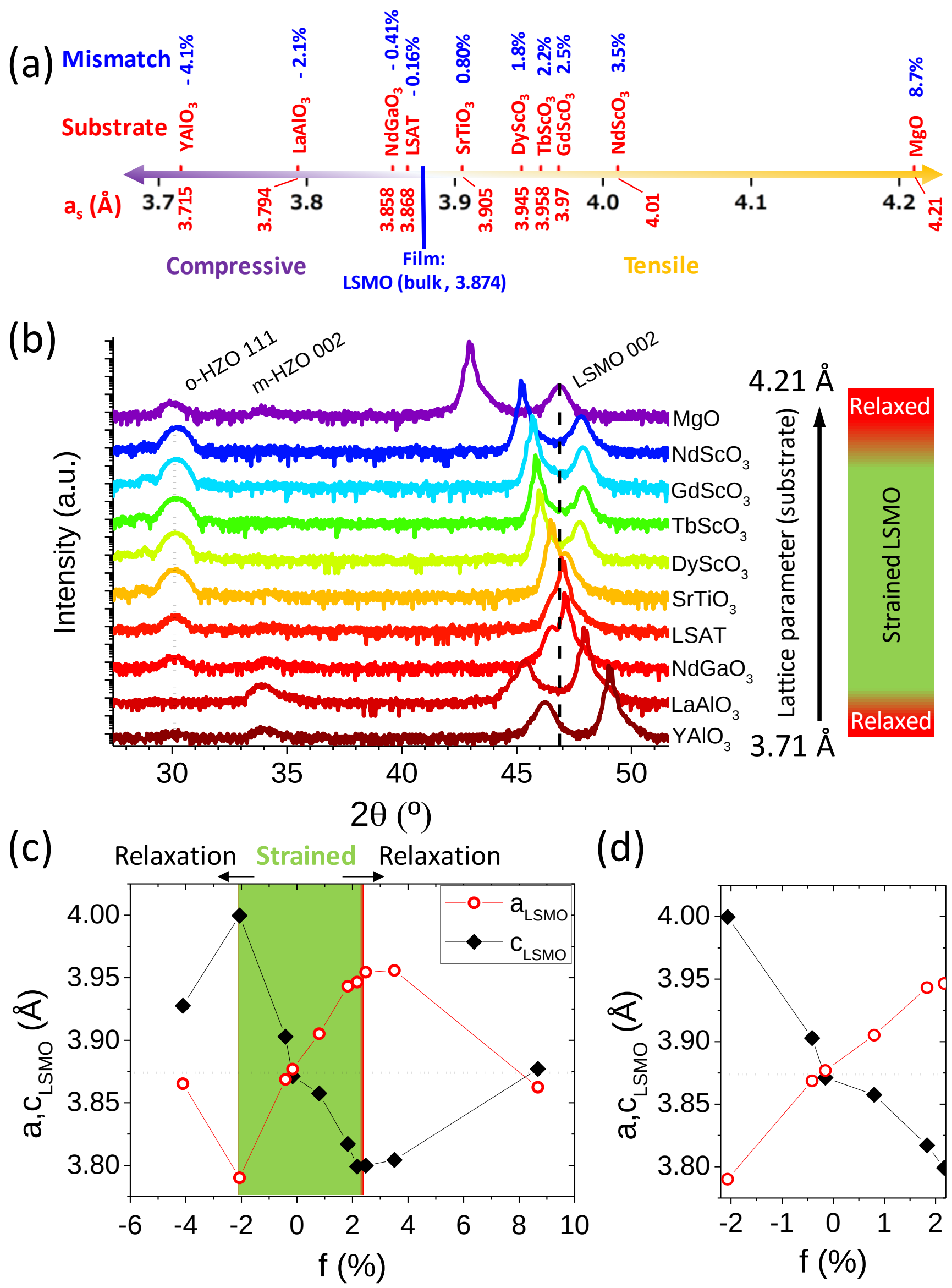

**Figure 1**. (a) Sketch showing the lattice mismatch between the LSMO electrode and the substrates used to deposit HZO/LSMO bilayers. The lattice mismatch (f) between LSMO and the substrates, defined as f = 100x($a_s$-$a_{LSMO}$)/$a_{LSMO}$, where $a_{LSMO}$ and as are the lattice parameters of bulk LSMO and the substrate, respectively. Pseudocubic cell is used for rhombohedral $LaAlO_3$ and orthorhombic ($NdGaO_3$ and scandates) substrates. (b) XRD θ-2θ symmetric scans of the HZO/LSMO bilayers. Scans are shifted vertically according the lattice parameter of the substrate (see labels and arrow at the right). Vertical solid line at 2θ = 30.1° marks the positions of the o-HZO 111 peak in the film on $SrTiO_3$(001). The vertical dashed line marks the position of the (002) reflection in bulk LSMO. Right: schematics of the strain state of LSMO depending on the lattice parameter of the substrate. (c) Out-of-plane and in-plane lattice parameters of LSMO as a function of the lattice mismatch with the substrate. (d) Zoom around the range of lattice mismatch where LSMO is elastically strained.

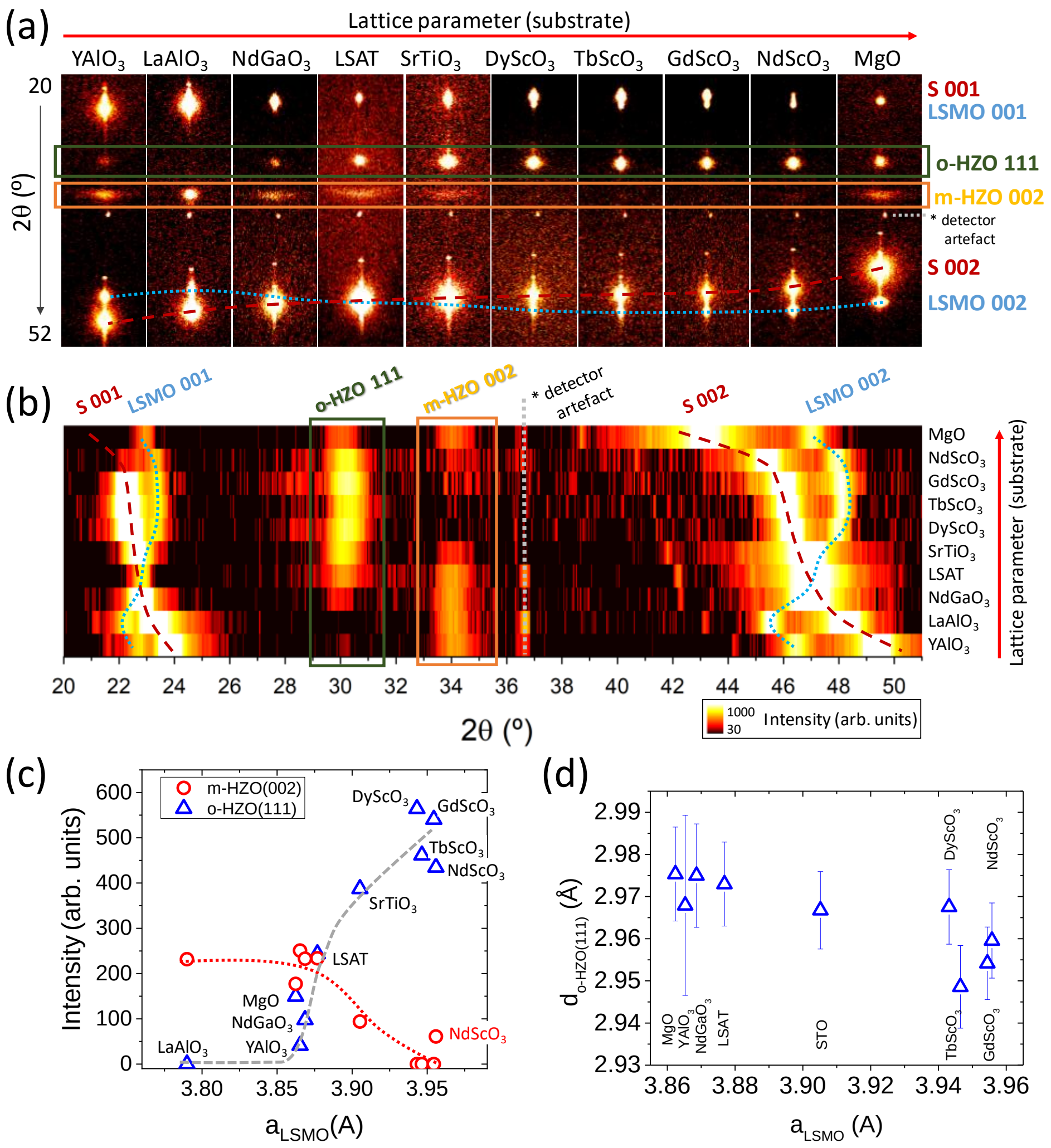

**Figure 2.** (a) XRD 2θ-χ frames of the HZO/LSMO bilayers. The 2θ and χ ranges are from 20° to 52° and from -8° to +8°, respectively. (b) Mapping of the orthorhombic and monoclinic phases as a function of the substrate lattice parameter. The 2θ scans were integrated from χ = -10° to +10° and the samples are ordered as the substrate lattice parameter increases. The change in the LSMO peaks position (marked with black dotted lines) on the used substrate (peaks position marked with red dashed lines) is also visualized. (c) Intensity of the o-HZO(111) and m-HZO(002) peaks (calculated from gaussian fits) and (d) interplanar $d_{o\text{-}HZO(111)}$ spacing as a function of the ip lattice parameter of the LSMO electrode. The $d_{o\text{-}HZO(111)}$ spacing was determined by Gaussian fits of the 2θ peak position, and the error bar is set to 1σ of the fit.

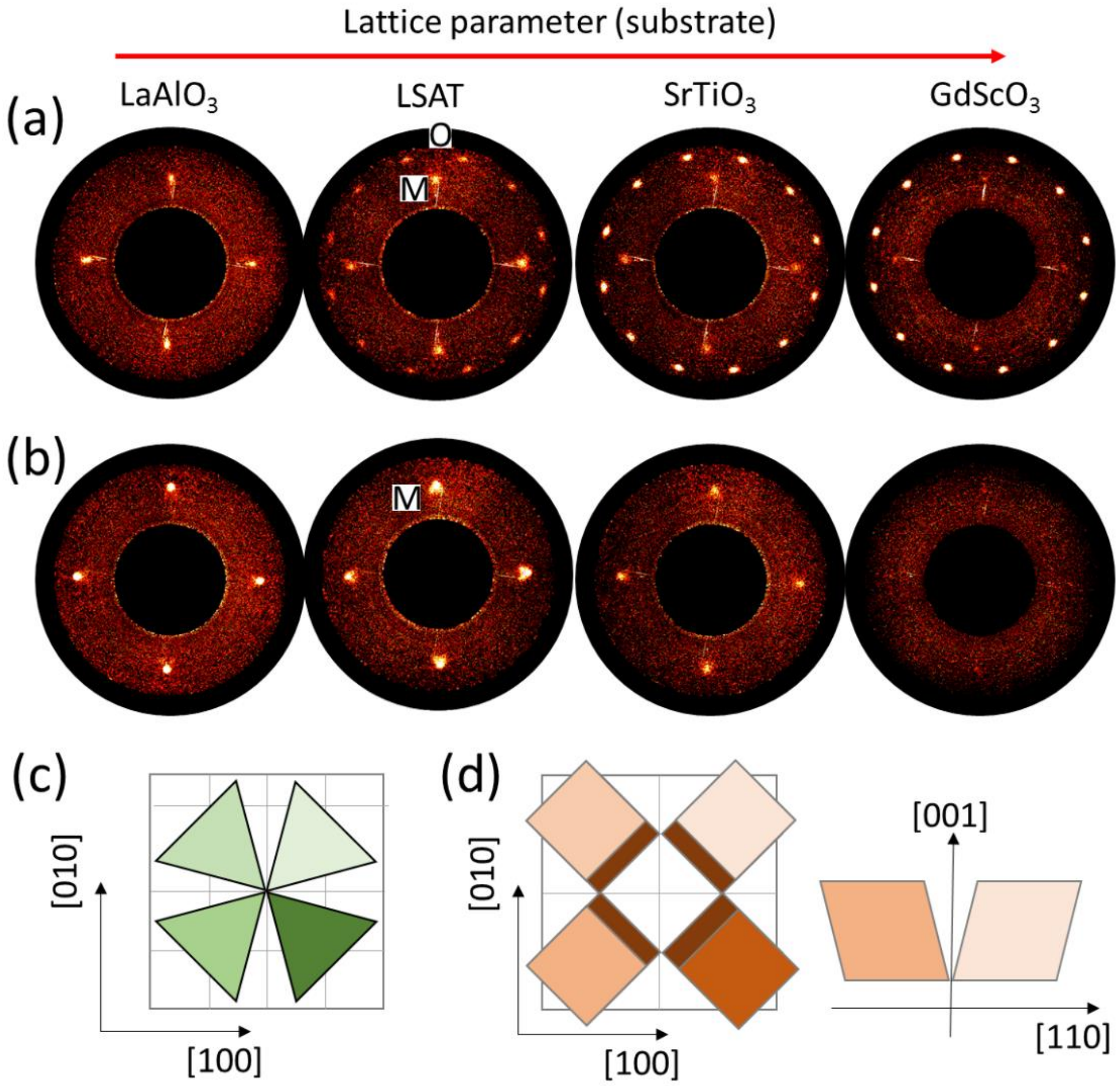

**Figure 3.** XRD pole figures of (a) o-HZO -111 (O) and m-HZO -111 (M), and (b) m-HZO -111 (M) reflections obtained for films on $LaAlO_3$, LSAT, $SrTiO_3$, and $GdScO_3$. The pole figures were measured in the range of $\chi$ from 35 to 80º. (c) Sketch of the epitaxial relationship of the o-HZO phase (top view). (d) Top and cross-sectional views of the epitaxial relationship of the m-HZO phase.

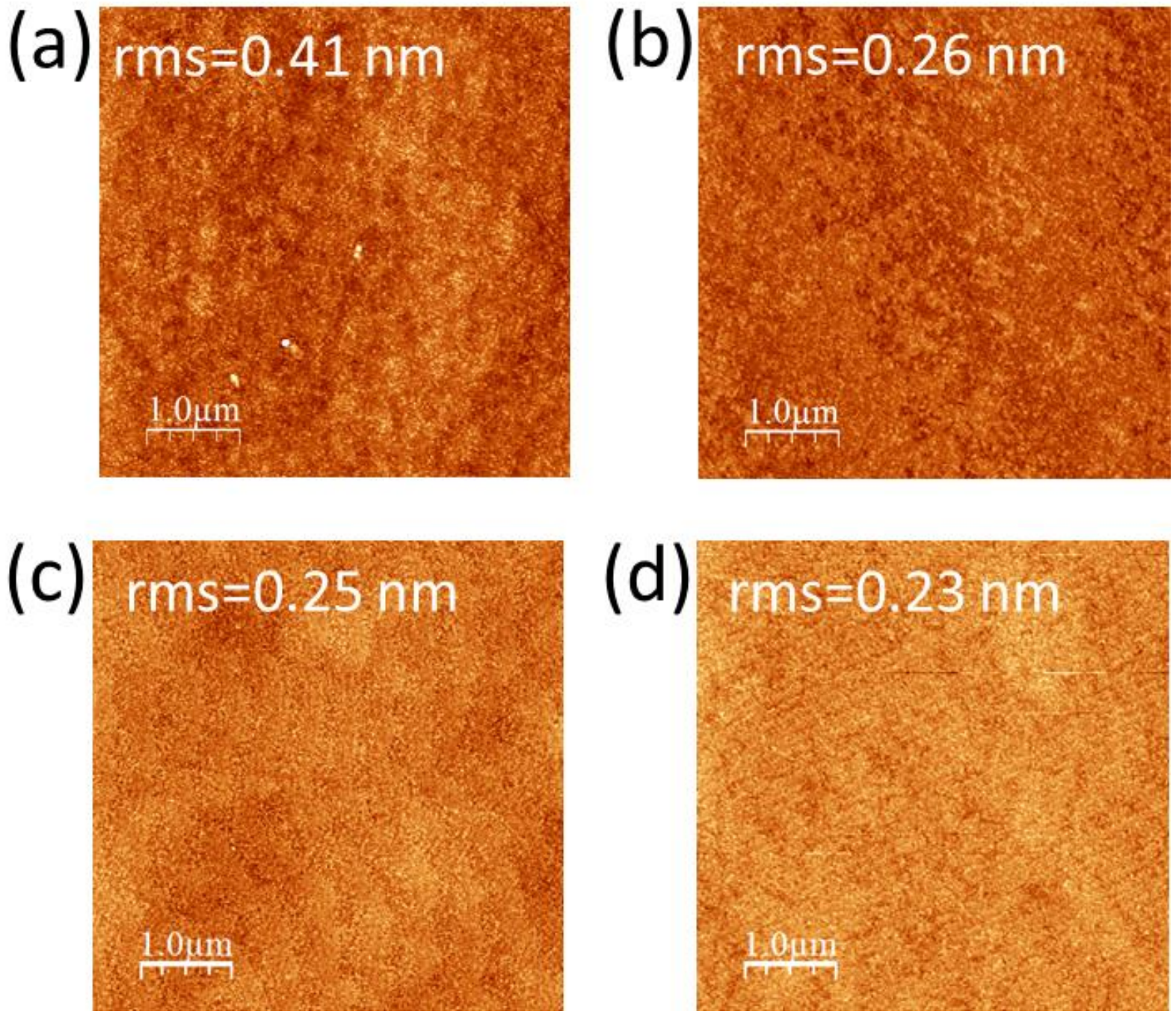

**Figure 4.** Topographic AFM images (5 µm x 5 µm) of HZO films on $NdScO_3$ (a), $DyScO_3$ (b), LSAT (c), and $LaAlO_3$ (d). The rms roughness is indicated in the top of each AFM image.

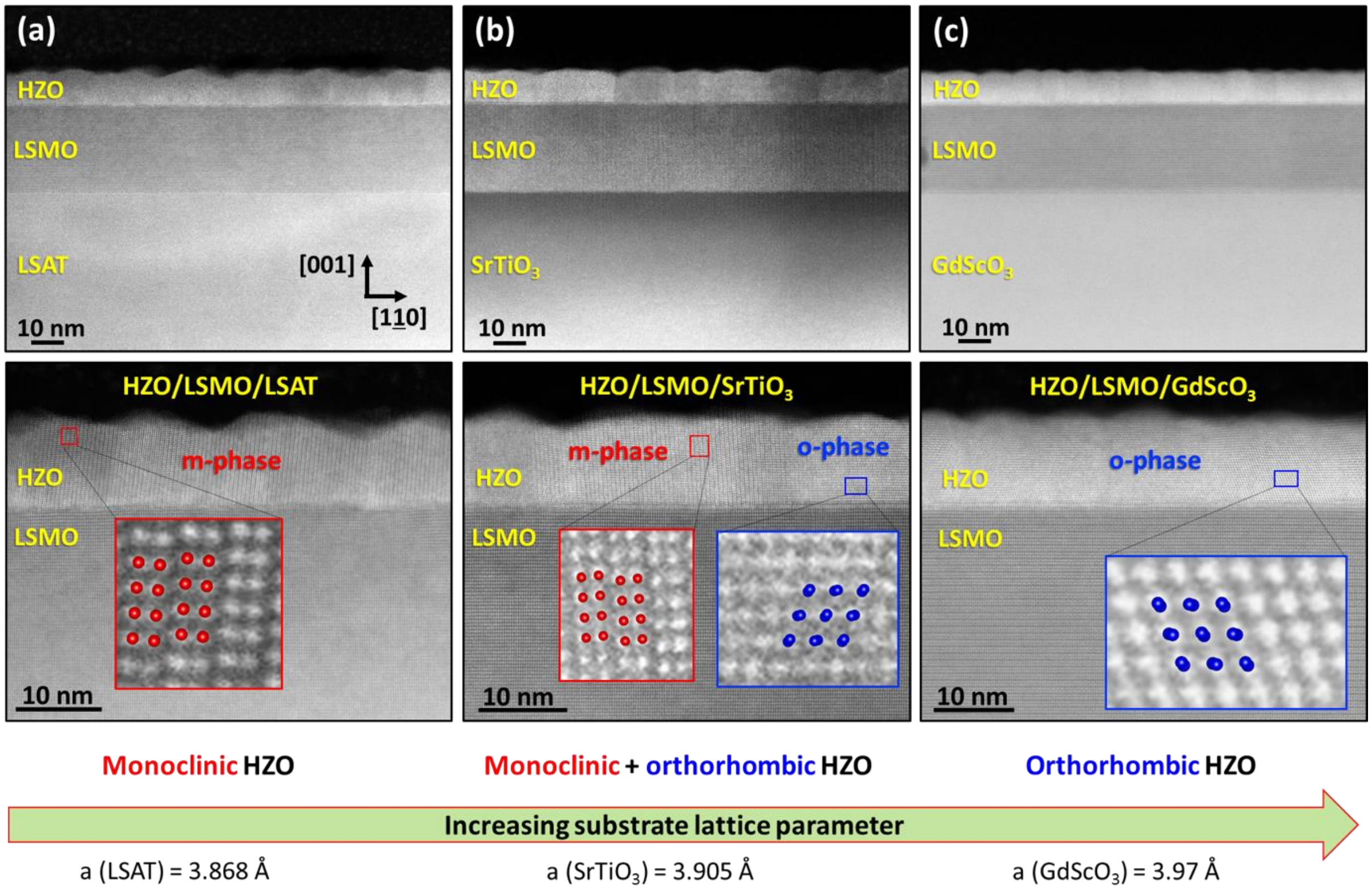

**Figure 5.** Cross-sectional HAADF STEM images of HZO/LSMO films on (a) LSAT, (b) $SrTiO_3$, and (c) $GdScO_3$ substrates. The images were acquired along the [110] zone axes of the substrates. Top panels are low magnification images showing the substrate, the LSMO and the HZO films. Bottom panels are higher magnification images of the HZO and LSMO films. The insets show atomic-resolution images of the HZO films. Red and blue circles depict the monoclinic (space group $P2_1/c$) and orthorhombic (space group $Pca2_1$) structures, respectively.

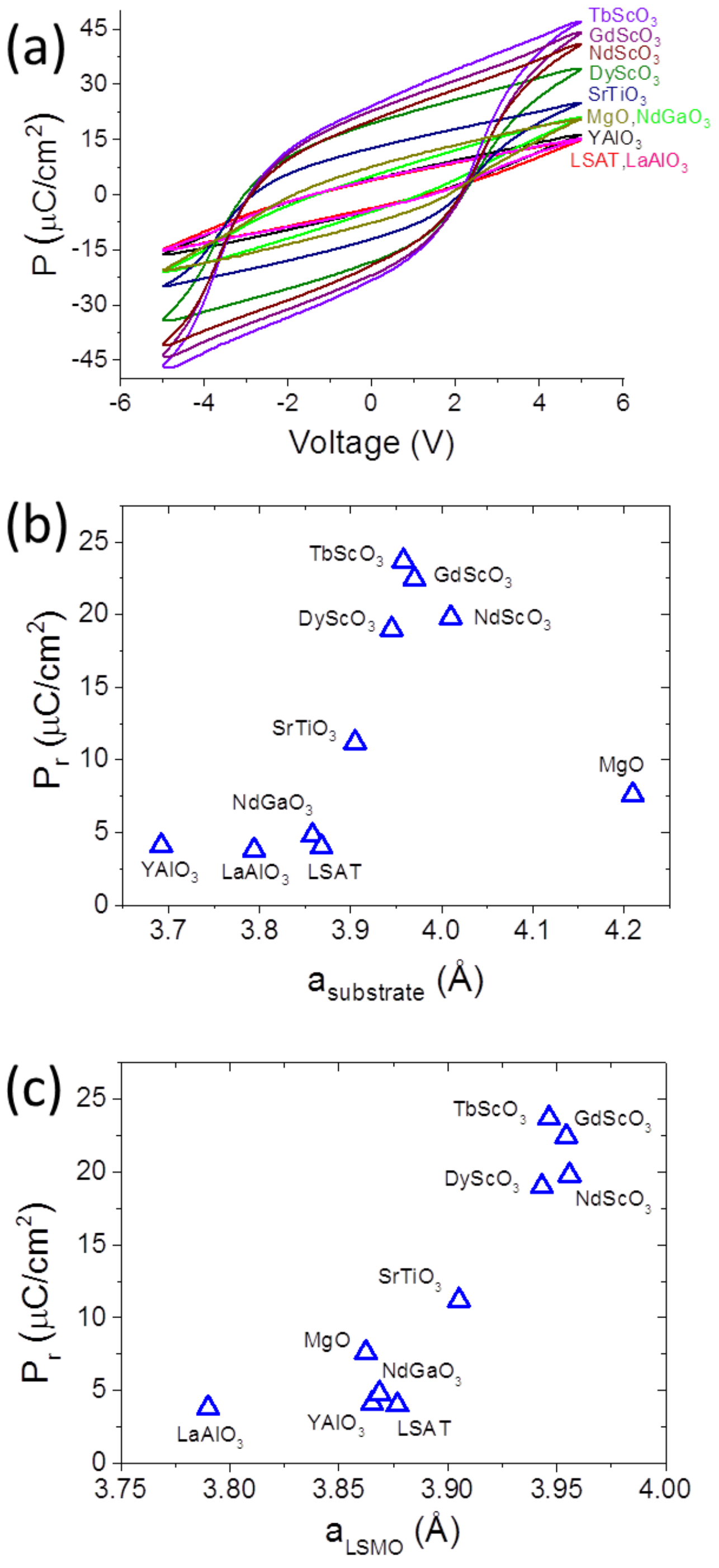

**Figure 6.** (a) Ferroelectric polarization loops of the HZO films. Remnant polarization as a function of (b) the lattice parameter of the substrate and (c) the ip parameter of the LSMO electrode.

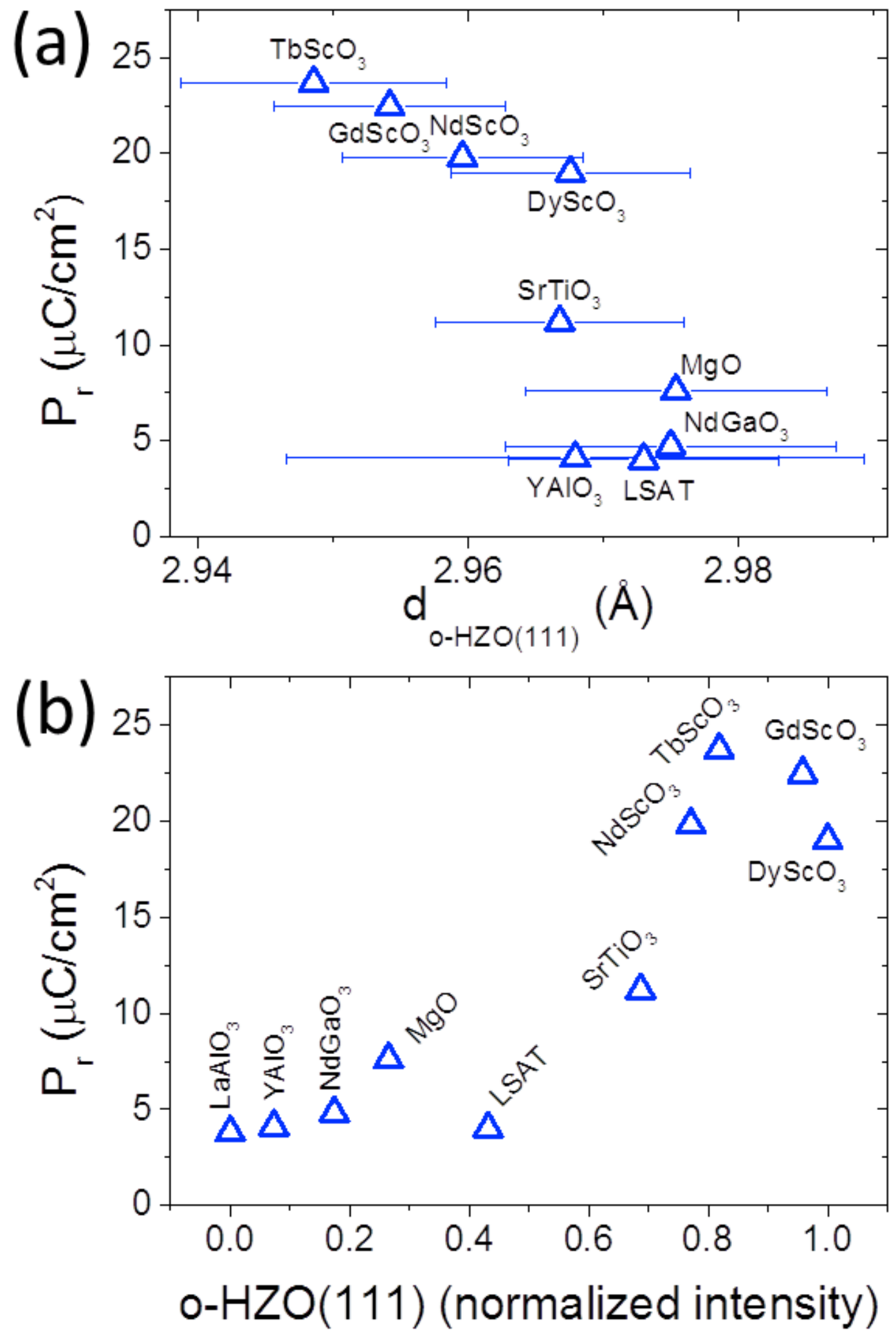

**Figure 7.** (a) Remnant polarization as a function of the interplanar $d_{o\text{-}HZO(111)}$ spacing. $d_{o\text{-}HZO(111)}$ was determined by Gaussian fits of the XRD 2θ peak position, and the error bar is set to 1σ of the fit. (b) Remnant polarization as a function of the normalized intensity of the XRD o-HZO 111 reflection.

REFERENCES

(1) Böscke, T. S.; Müller, J.; Bräuhaus, D.; Schröder, U.; Böttger, U. Ferroelectricity in Hafnium Oxide Thin Films. *Appl. Phys. Lett.* **2011**, 99, 102903.

(2) Park, M. H.; Lee, Y. H.; Kim, H. J.; Kim, Y. J.; Moon, T.; Kim, K. D.; Müller, J.; Kersch, A.; Schroeder, U.; Mikolajick, T.; Hwang, C. S. Ferroelectricity and Antiferroelectricity of Doped Thin $HfO_2$-Based Films. *Adv. Mater.* **2015**, 27, 1811−1831.

(3) Mikolajick, T.; Slesazeck, S.; Park, M. H.; Schroeder, U. Ferroelectric Hafnium Oxide for Ferroelectric Random-Access Memories and Ferroelectric Field-Effect Transistors. *MRS Bull.* **2018**, 43, 340.

(4) Park, M. H.; Lee, Y. H.; Mikolajick, T.; Schroeder, U.; Hwang C. S. Review and Perspective on Ferroelectric $HfO_2$-based Thin Films for Memory Applications. *MRS Comm.* **2018**, 8, 795.

(5) Hoffmann, M.; Schroeder, U.; Schenk, T.; Shimizu, T.; Funakubo, H.; Sakata, O.; Pohl, D.; Drescher, M.; Adelmann, C.; Materlik, R.; Kersch, A.; Mikolajick, T. Stabilizing the Ferroelectric Phase in Doped Hafnium Oxide. *J. Appl. Phys.* **2015**, 118, 072006.

(6) Lomenzo, P. D.; Chung, C. C.; Zhou, C. Z.; Jones, J. L.; Nishida, T. Doped $Hf_{0.5}Zr_{0.5}O_2$ for High Efficiency Integrated Supercapacitors. *Appl. Phys. Lett.* **2017**, 23, 232904.

(7) Kim, S. J.; Mohan, J.; Lee, J. S.; Kim, Harrison S.; Lee, J.; Young, C. D.; Colombo, L.; Summerfelt, S. R.; San, T.; Kim, J. Stress-Induced Crystallization of Thin $Hf_{1-x}Zr_xO_2$ Films: The Origin of Enhanced Energy Density with Minimized Energy Loss for Lead-Free Electrostatic Energy Storage Applications. *ACS Appl. Mater. Interf.* **2019**, 11, 5208-5214.

(8) Lee, Y. H.; Hyun, S. D.; Kim, H. J.; Kim, J. S.: Yoo, C.; Moon, T.; Kim, K. D.; Park, H. W.; Lee, Y. B.; Kim, B. S.; Roh, J.; Park, M. H.; Hwang, C. S. Nucleation-Limited Ferroelectric Orthorhombic Phase Formation in $Hf_{0.5}Zr_{0.5}O_2$ Thin Films. *Adv. Electron. Mater.* **2019**, 5, 1800436.

(9) O'Connor, E.; Halter, M.; Eltes, F.; Sousa, M.; Kellock, A.; Abel, S.; Fompeyrine, J. Stabilization of Ferroelectric $Hf_xZr_{1-x}O_2$ Films using a Millisecond Flash Lamp Annealing Technique. *APL Mater.* **2019**, 6, 121103.

(10) Migita, S.; Ota, H.; Shibuya, K.; Yamada, H.; Sawa, A.; Matsukawa, T.; Toriumi, A. Phase Transformation Behavior of Ultrathin $Hf_{0.5}Zr_{0.5}O_2$ Films Investigated through Wide Range Annealing Experiments. *Jpn. J. Appl. Phys.* **2019**, 58, SBBA07.

(11) Kim, S.J.; Mohan, J.; Summerfelt, S.R.; Kim, J. Ferroelectric $Hf_{0.5}Zr_{0.5}O_2$ Thin Films: A Review of Recent Advances. *JOM* **2019**, 71, 246.

(12) Park, M. H.; Lee, Y. H.; Mikolajick, T.; Schroeder, S.; Hwang, C. C. Thermodynamic and Kinetic Origins of Ferroelectricity in Fluorite Structure Oxides. *Adv. Electron. Mater.* **2019**, 5, 1800522.

(13) Katayama, K.; Shimizu, T.; Sakata, O.; Shiraishi, T.; Nakamura, S.; Kiguchi, T.; Akama, A.; Konno, T. J.; Uchida, H.; Funakubo, H. Orientation Control and Domain Structure Analysis of {100}-Oriented Epitaxial Ferroelectric Orthorhombic $HfO_2$-based Thin Films. *J. Appl. Phys.* **2016**, 119, 134101.

(14) Mimura, T.; Shimizu, T.; Uchida, H.; Sakata, O.; Funakubo, H. Thickness-Dependent Crystal Structure and Electric Properties of Epitaxial Ferroelectric $Y_2O_3$-$HfO_2$ Films. *Appl. Phys. Lett.* **2018**, 113, 102901.

(15) Li, T.; Zhang, N.; Sun, Z.; Xie, C.; Ye, M.; Mazumdar, S.; Shu, L.; Wang, Y.; Wang, D.; Chen, L.; Ke, S.; Huang, H. Epitaxial Ferroelectric $Hf_{0.5}Zr_{0.5}O_2$ Thin Film on a Buffered YSZ Substrate through Interface Reaction. *J. Mater. Chem. C* **2018**, 6, 9224.

(16) Shimizu, T.; Katayama, K.; Kiguchi, T.; Akama, A.; Konno, T. J.; Sakata, O.; Funakubo, H. The Demonstration of Significant Ferroelectricity in Epitaxial Y-doped $HfO_2$ Film. *Sci. Rep.* **2016**, 6, 32931.

(17) Yoong, H. Y.; Wu, H.; Zhao, J.; Wang, H.; Guo, R.; Xiao, J.; Zhang, B.; Yang, P.; Pennycook, S. J.; Deng, N.; Yan, X.; Chen, J. Epitaxial Ferroelectric $Hf_{0.5}Zr_{0.5}O_2$ Thin Films and Their Implementations in Memristors for Brain Inspired Computing. *Adv. Funct. Mater.* **2018**, 28, 1806037.

(18) Lyu, J.; Fina, I.; Solanas, R.; Fontcuberta, J.; Sánchez, F. Robust Ferroelectricity in Epitaxial $Hf_{1/2}Zr_{1/2}O_2$ Thin Films. *Appl. Phys. Lett.* **2018**, 113, 082902.

(19) Lyu, J.; Fina, I.; Solanas, R.; Fontcuberta, J.; Sánchez. Growth Window of Ferroelectric Epitaxial $Hf_{0.5}Zr_{0.5}O_2$ Thin Films. *ACS Appl. Electron. Mater.* **2019**, 1, 220.

(20) Wei, Y.; Nukala, P.; Salverda, M.; Matzen, S.; Zhao, H. J.; Momand, J.; Everhardt, A.; Agnus, G.; Blake, G. R.; Lecoeur, P.; Kooi, B. J.; Íñiguez, J.; Dkhil, B.; Noheda, B. A Rhombohedral Ferroelectric Phase in Epitaxially-Strained $Hf_{0.5}Zr_{0.5}O_2$ Thin Films. *Nat. Mater.* **2018**, 17, 1095.

(21) Li, T.; Ye, M.; Sun, Z.; Zhang, N.; Zhang, W.; Inguva, S.; Xie, C.; Chen, L.; Wang, Y.; Ke, S.; Huang, H. Origin of Ferroelectricity in Epitaxial Si-Doped $HfO_2$ Films. *ACS Appl. Mater. Interf.* **2019**, 11, 4139.

(22) Lyu, J.; Fina, I.; Fontcuberta, J.; Sánchez, F. Epitaxial Integration on Si(001) of Ferroelectric $Hf_{0.5}Zr_{0.5}O_2$ Capacitors with High Retention and Endurance. *ACS Appl. Mater. Interf.* **2019**, 11, 6224.

(23) Batra, R.; Huan, T. D.; Jones, J. L.; Rossetti, G.; Ramprasad, R. Factors Favoring Ferroelectricity in Hafnia: A First-Principles Computational Study. *J. Phys. Chem. C* **2017**, 121, 4139.

(24) Gorbenko, O. Y.; Samoilenkov, S. V.; Graboy, I. E.; Kaul, A. R. Epitaxial Stabilization of Oxides in Thin Films. *Chem. Mater.* **2002**, 14, 4026.

(25) Kaul, A. R.; Gorbenko, O. Y.; Kamenev, A. A. The Role of Heteroepitaxy in the Development of New Thin-Film Oxide-based Functional Materials. *Russian Chem. Rev.* **2004**, 73, 861.

(26) Dix, N.; Muralidharan, R.; Varela, M.; Fontcuberta, J.; Sánchez, F. Mapping of the Epitaxial Stabilization of Quasi-Tetragonal $BiFeO_3$ with Deposition Temperature. *Appl. Phys. Lett.* **2012**, 100, 122905.

(27) Gich, M.; Fina, I.; Morelli, A.; Sánchez, F.; Alexe, M.; Gàzquez, J.; Fontcuberta, J.; Roig, A. Multiferroic Iron Oxide Thin Films at Room Temperature. *Adv. Mater.* **2014**, 26, 4645.

(28) Materlik, R.; Künneth, C.; Kersch, A. The origin of ferroelectricity in $Hf_{1-x}Zr_xO_2$: A computational investigation and a surface energy model. *J. Appl. Phys.* **2015**, 117, 134109.

(29) Liu, S.; Hanrahan, B. M. Effects of growth orientations and epitaxial strains on phase stability of $HfO_2$ thin films. *Phys. Rev. Mater.* **2019**, 3, 054404.

(30) Schlom, D. G., Chen, L. Q.; Eom, C. B.; Rabe, K. M.; Streiffer, S. K.; Triscone, J.M. Strain Tuning of Ferroelectric Thin Films. *Annu. Rev. Mater. Res.* **2007**, 37, 589.

(31) Choi, K. J.; Biegalski, M.; Li, Y. L.; Sharan, A.; Schubert, J.; Uecker, R.; Reiche, P.; Chen, Y. B.; Pan, X. Q.; Gopalan, V.; Chen, L.Q.; Schlom, D. G.; Eom, C. B. Enhancement of Ferroelectricity in Strained $BaTiO_3$ Thin Films. *Science* **2004**, 306, 1005.

(32) Batra, R.; Tran, H. D.; Ramprasad, R. Stabilization of metastable phases in hafnia owing to surface energy effects. *Appl. Phys. Lett.* **2016**, 108, 172902.

(33) Grundmann, M; Böntgen, T.; Lorenz, M. Occurrence of Rotation Domains in Heteroepitaxy. *Phys. Rev. Lett.* **2010**, 105, 146102.

(34) Scigaj, M.; Dix, N.; Cabero, M.; Rivera-Calzada, A.; Santamaría, J.; Fontcuberta, J.; Herranz, G.; Sánchez, F. Yttria-Stabilized Zirconia/$SrTiO_3$ Oxide Heteroepitaxial Interface with Symmetry Discontinuity. *Appl. Phys. Lett.* **2014**, 104, 251602.

(35) Zhou, H.; Wang, H. Q.; Li, Y.; Li, K.; Kang, J.; Zheng, J. C.; Jiang, Z.; Huang, Y.; Wu, L.; Zhang, L.; Kisslinger, K.; Zhu, Y. Evolution of Wurtzite ZnO Films on Cubic MgO (001) Substrates: A Structural, Optical, and Electronic Investigation of the Misfit Structures. *ACS Appl. Mater. Interf.* **2014**, 6, 13823.

(36) Trampert, A; Ploog, K. H. Heteroepitaxy of Large-Misfit Systems: Role of Coincidence Lattice. *Cryst. Res. Technol.* **2000**, 35, 793.

(37) Sánchez, F.; Bachelet, R.; de Coux, P.; Warot-Fonrose, B.; Skumryev, V.; Tarnawska, L.; Zaumseil, P.; Schroeder, T.; Fontcuberta, J.; Domain Matching Epitaxy of Ferrimagnetic $CoFe_2O_4$ Thin Films on $Sc_2O_3$/Si(111). *Appl. Phys. Lett.* **2011**, 99, 211910.

(38) Venzke, S.; van Dover, R. B.; Phillips, J. M.; Gyorgy, E. M.; Siegrist, T.; Chen, C. H.; Werder, D.; Fleming, R. M.; Felder, R. J.; Coleman, E.; Opila, R. Epitaxial Growth and Magnetic Behavior of $NiFe_2O_4$ Thin Films. *J. Mater. Res.* **1996**, 11, 1187.

(39) Meyer, R.; Waser, R.; Prume, K.; Schmitz, T.; Tiedke, S. Dynamic Leakage Current Compensation in Ferroelectric Thin-Film Capacitor Structures. *Appl. Phys. Lett.* **2005**, 86, 142907.

(40) Fina, I.; Fàbrega, L.; Langenberg, E.; Martí, X.; Sánchez, F.; Varela, M.; Fontcuberta, J. Non-Ferroelectric Contributions to the Hysteresis Cycles in Manganite Thin Films: a Comparative Study of Measurement Techniques. *J. Appl. Phys.* **2011**, 109, 074105.

Supporting Information

# Engineering Ferroelectric $Hf_{0.5}Zr_{0.5}O_2$ Thin Films by Epitaxial Stress

Saúl Estandía,[1] Nico Dix,[1] Jaume Gazquez,[1] Ignasi Fina,[1] Jike Lyu,[1] Matthew F. Chisholm,[2] Josep Fontcuberta,[1] Florencio Sánchez[1,*]

[1] Institut de Ciència de Materials de Barcelona (ICMAB-CSIC), Campus UAB, Bellaterra 08193, Barcelona, Spain

[2] Center for Nanophase Materials Sciences, Oak Ridge National Laboratory, Tennessee 37831-6064, USA

**XRD reciprocal space maps around asymmetric LSMO reflections.**

Reciprocal space maps around pseudocubic 103 substrate and LSMO 103 film reflections (the only exception is the RSM measured for MgO which due to systematic extinction of the 103 reflection, and thus was measured around the 113 cubic substrate reflection – thus the in-plane parameter is $a_{100}=a_{110}/\sqrt{2}$). Vertical lines indicate the in-plane position of the substrate peaks.

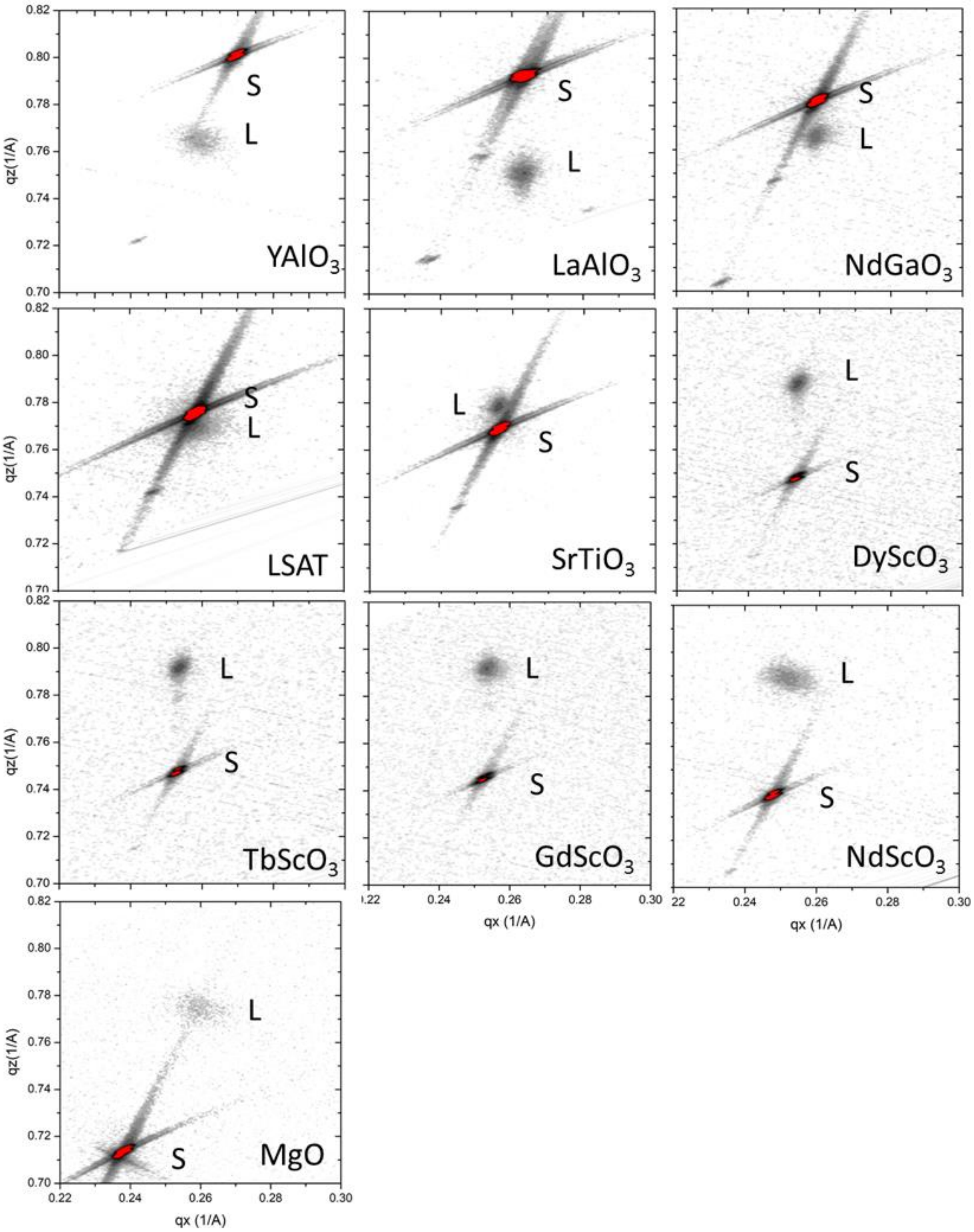

**Figure S1**: RSM around 103 pseudocubic substrate reflection (Note: the sample on MgO substrate is measured at 113 reflection due to absence of 103 reflection, and thus the x scale of MgO 113 is divided by √2, to allow easy comparison to 103 reflections. The labels L and S correspond to LSMO and substrate reflections, respectively.

**Simulation of Laue fringes around o-HZO 111 reflection.**

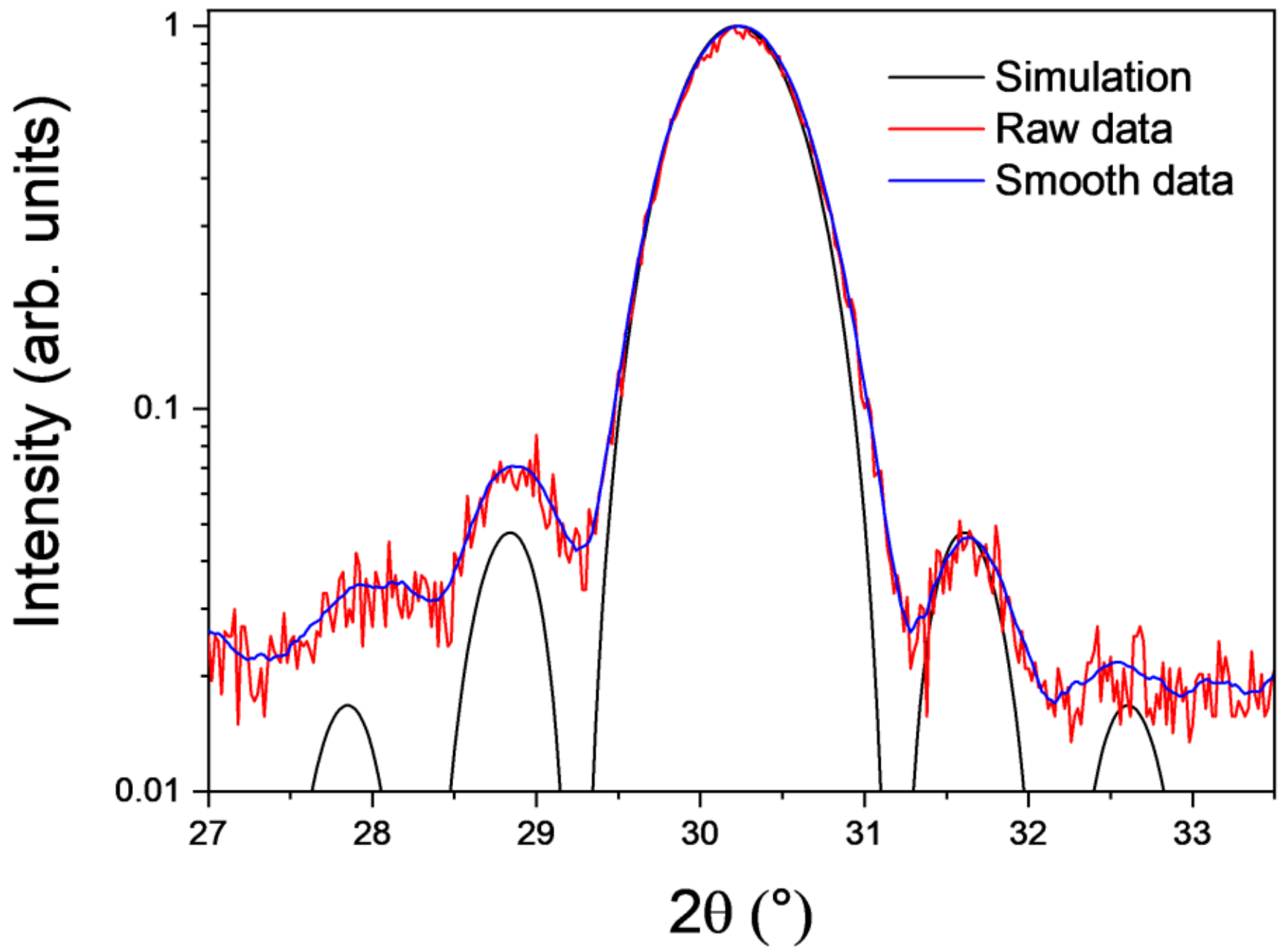

**Figure S2**: XRD (Cu Kα radiation) θ-2θ symmetric scan (red curve) around the o-HZO 111 reflection of the HZO/LSMO/$NdScO_3$(001) sample. The blue curve corresponds to the experimental data after applying a smoothing filter. The peak at around 2θ = 30.2° corresponds to the o-HZO 111 reflection, and the peaks at both sides are Laue reflections. The data has been simulated (black curve) according the dependence:[1]

$$I(Q) = \left( \frac{\sin\left(\frac{QNc}{2}\right)}{\sin\left(\frac{Qc}{2}\right)} \right)^2$$

where Q = 4πsin(θ)/λ is the reciprocal space vector, N the number of unit cells along the out-of-plane direction and c the corresponding interplanar spacing. The simulation is fitted to the experimental curve supposing a thickness of 9.46 Å (N = 32 and c = 2.957 Å) and being the corresponding Bragg reflection located at 2θ = 30.22°.

**XRD reciprocal space maps around asymmetric HZO**

Reciprocal space maps (RSM) around HZO 331 and HZO 402 are shown in Figure S3 (a) and (c) for $SrTiO_3$ and in Figure S3 (b) and (d) for $GdScO_3$ substrates, respectively. The HZO 331 reflections are in proximity to pseudocubic (pc) substrate 113 and LSMO 113 film reflections, while the HZO 402, and pseudocubic LSMO 1.5 1.5 2.5 and $GdScO_3$ 1.5 1.5 2.5 reflections are close ($SrTiO_3$ does not show 1.5 1.5 2.5 diffraction peak due to cubic crystal symmetry). Considering the epitaxial relationships [1-10]o-HZO(111)//[1-10]LSMO(001)//[1-10]Substrate(001) and [11-2]o-HZO(111)//[1-10]LSMO(001)//[1-10]Substrate(001). The $q_z$ and $q_x$ axes of the RSM around HZO 331 correspond to HZO[111] and HZO[11-2] directions, while RSM around the HZO 402 reflection relate to HZO[111] and HZO[1-10] directions. The in-plane parameters for HZO indicate a cell diagonal of 6.27±0.01Å (6.25±0.01Å) along [11-2]HZO(111) and of 7.15±0.01Å (7.21±0.01Å) along [110]HZO(111) for LSMO/$SrTiO_3$ (LSMO/$GdScO_3$). The out-of-plane (111) spacing is 2.96±0.01Åfor LSMO/$SrTiO_3$ and 2.95±0.01Å for LSMO/$GdScO_3$.

The calculated lattice constants for bulk orthorhombic $HfO_2$ ($P_{ca}2_1$) are a = 5.234 Å, b = 5.010 Å and c=5.0431 Å.[2] The lattice constants of o-$Hf_{0.5}Zr_{0.5}O_2$ film will be very close as the atomic radii of Hf (1.44 Å) and Zr (1.45 Å) are very similar.[3] Thus, the cell diagonals for relaxed HZO will be around 6.24 Å along [11-2]HZO(111) and 7.21 Å along [1-10]HZO(111), in good agreement with the values of our films.

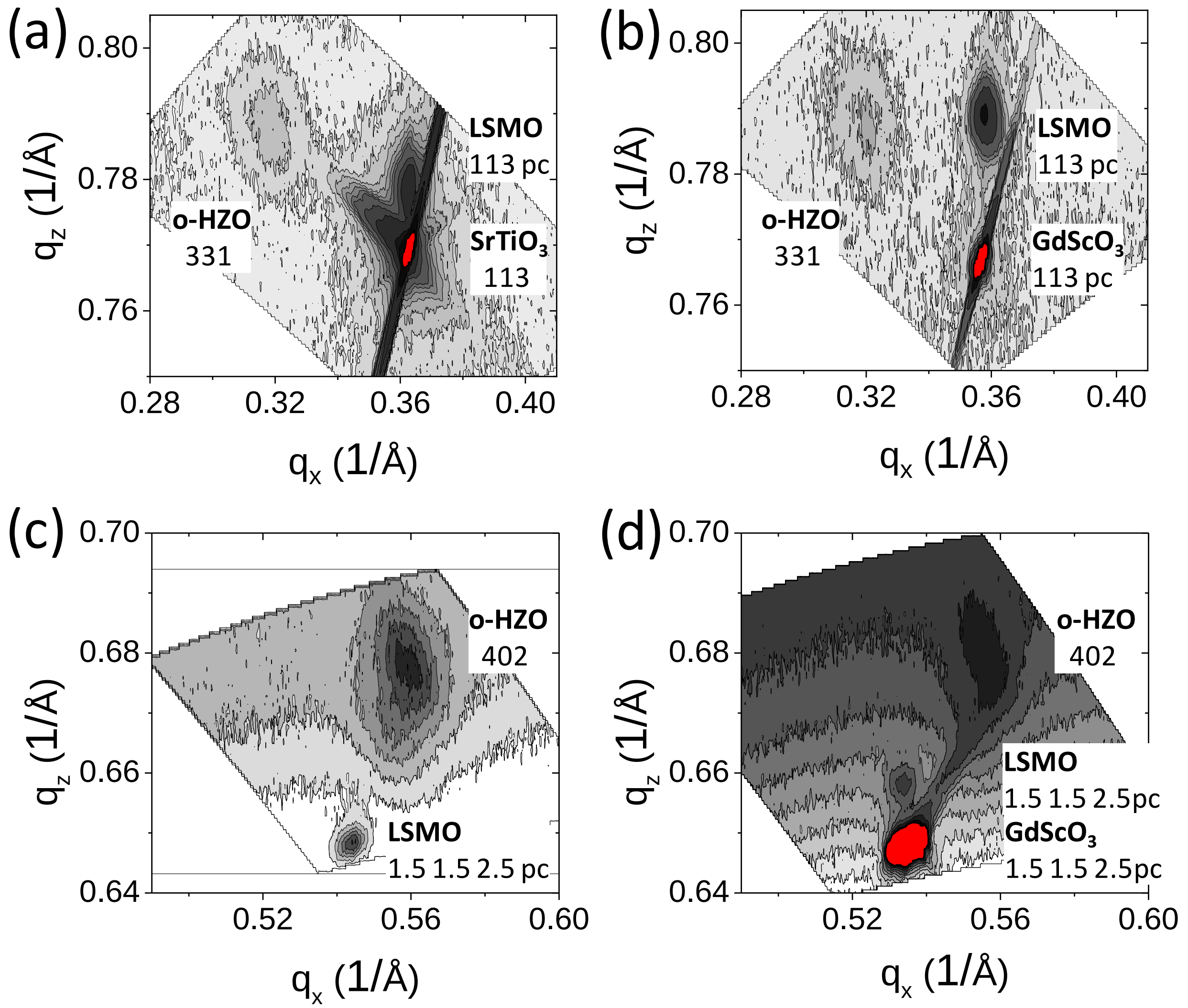

**Figure S3**: Reciprocal space maps around HZO 331 for the films on (a) LSMO/$SrTiO_3$ and (b) LSMO/$GdScO_3$. The corresponding maps around HZO 402 for the same samples are shown in (c) and (d), respectively.

**XRD ϕ-scans around o-HZO, m-HZO and substrates reflections**

ϕ-scans around asymmetrical reflections of the orthorhombic and monoclinic phases, derived from the XRD pole figures corresponding to films on $LaAlO_3$, LSAT, $SrTiO_3$ and $GdScO_3$ substrates shown in Figure 3, are presented in Figure S4. The asymmetrical reflections are 200 and 11-1 for the o-HZO phase, and 111 and 11-1 for the m-HZO phase.

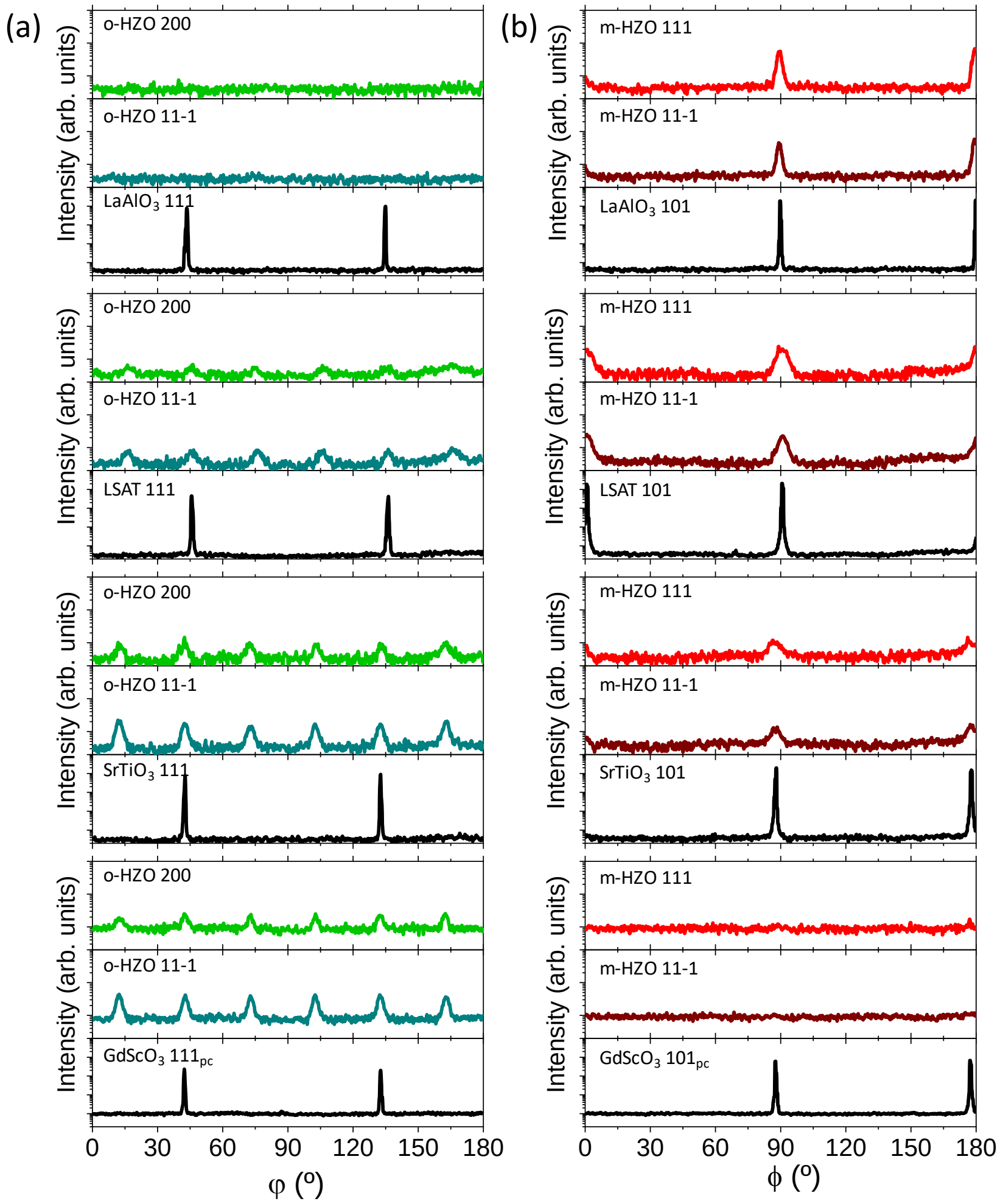

**Figure S4**: (a) XRD ϕ-scans around 200 and 11-1 reflections of o-HZO and 111 pseudocubic reflections of the substrates. In (b) ϕ-scans of 111 and 11-1 reflections of m-HZO and 101 pseudocubic reflections of the substrates. Intensity is plotted in logarithmic scale.

**Topographic images (atomic force microscopy) of all films.**

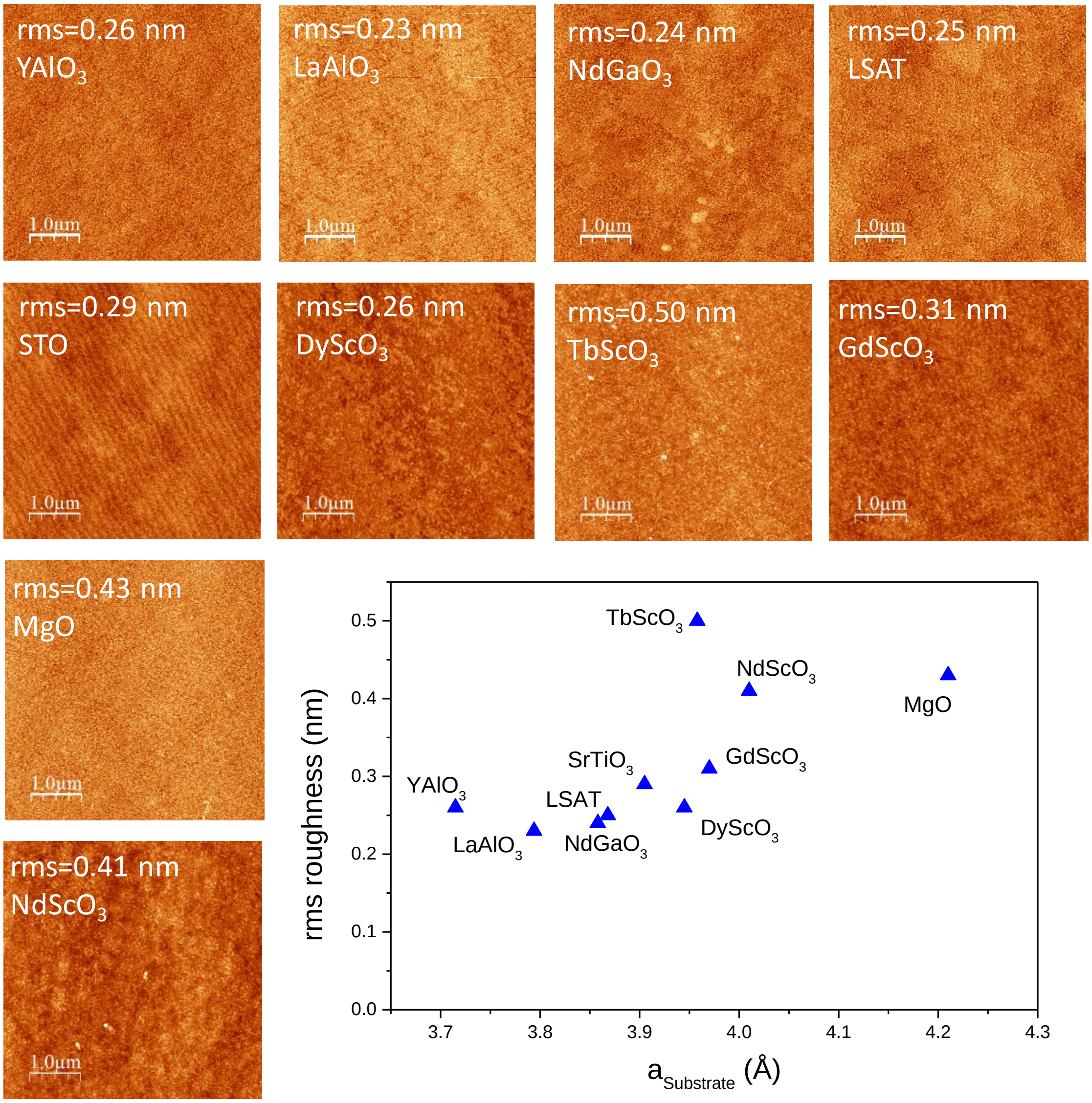

**Figure S5**: Topographic AFM images, 5 µm x 5 µm scanned area, of all HZO/LSMO/substrate samples. The substrate and the rms roughness is indicated in the top left of each image. The graph shows the rms roughness of the HZO films plotted as a function of the substrate lattice parameter.

## Ferroelectric measurements.

Ferroelectric measurements were performed with an AixACCT TFAnalyser2000 in the top-bottom electrode configuration at 1000 Hz. The area of the contacts was 283 µm2 for all the samples. Leakage compensation was performed by means of Dynamic leakage current compensation technique (DLCC).[4] Residual leakage exponential contribution (after DLCC) was first fitted and after removed from the measured current.

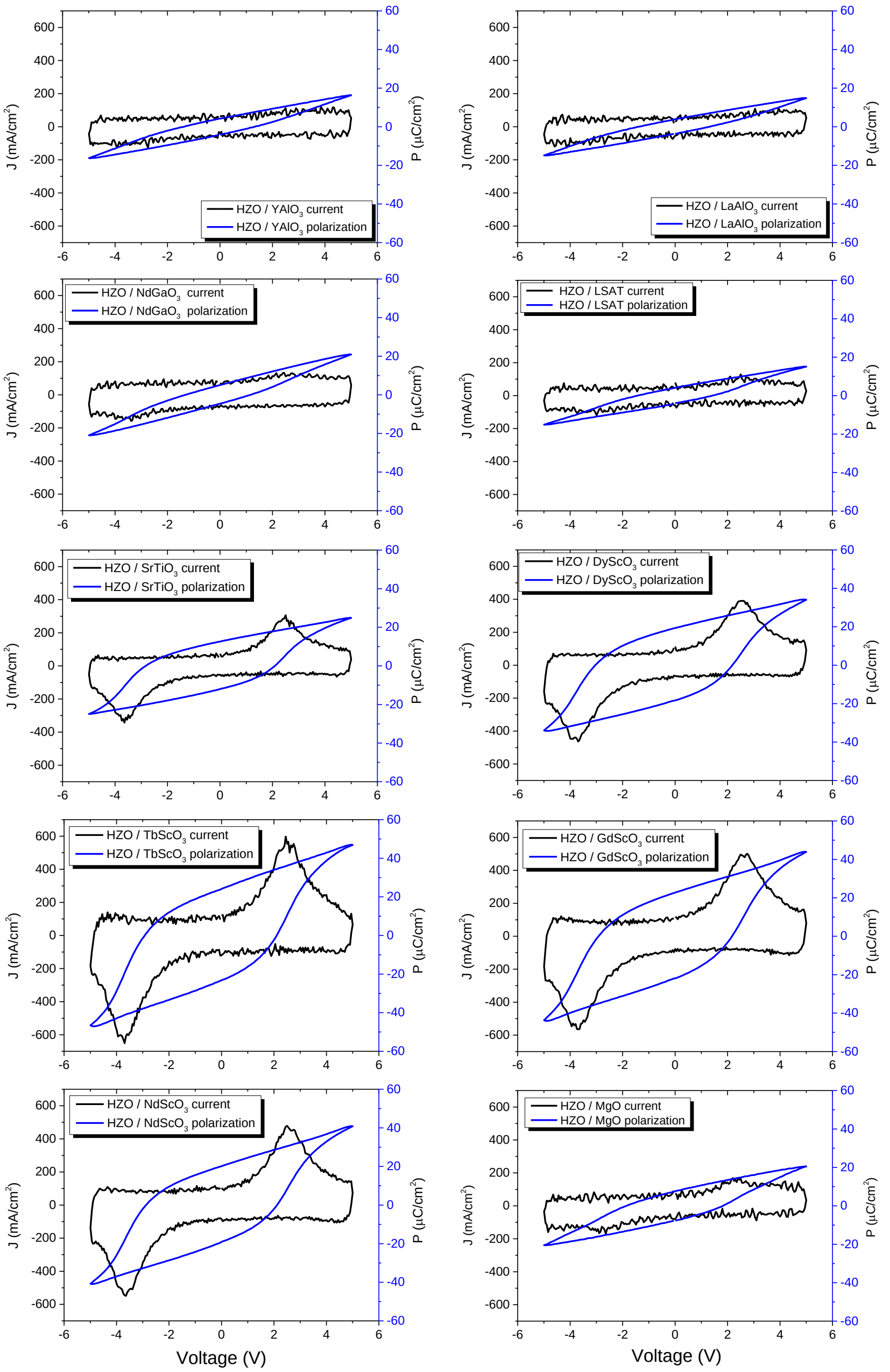

**Figure S6**: Ferroelectric polarization loops (in blue) and current-voltage curves (in black) of all the HZO films. The substrate is indicated in the top left of each panel.

## References

(1) Pesquera, D.; Martí, X.; Holy, V.; Bachelet, R.; Herranz, G.; Fontcuberta, J. X-Ray Interference Effects on the Determination of Structural Data in Ultrathin $La_{2/3}Sr_{1/3}MnO_3$ Epitaxial Thin Films. *Appl. Phys. Lett.* **2011**, 99, 221901.

(2) Tao, L.L.; Paudel, T.R.; Kovalev, A.A.; Tsymbal, E.Y. Reversible Spin Texture in Ferroelectric $HfO_2$. *Phys. Rev. B* **2017**, 95, 245141.

(3) Rignanese, G.M.; Gonze, X.; Jun, G.; Cho, K.; Pasquarello, A. First-Principles Investigation of High-κ Dielectrics: Comparison between the Silicates and Oxides of Hafnium and Zirconium. *Phys. Rev. B* **2004**, 69, 184301.

(4) Meyer, R.; Waser, R.; Prume, K.; Schmitz, T.; Tiedke, S. Dynamic Leakage Current Compensation in Ferroelectric Thin-Film Capacitor Structures. *Appl. Phys. Lett.* **2005**, 86, 142907.